\documentclass[aps,prb,preprint,unsortedaddress,showpacs]{revtex4}
\usepackage{amssymb}
\usepackage{graphicx}
\usepackage{color}

\begin{document}

\title{Electronic, vibrational, and thermodynamic properties of ZnS (zincblende and rocksalt structure)}

\author{M. Cardona}
\author{R. K. Kremer}
\email[Corresponding author:~E-mail~]{r.kremer@fkf.mpg.de}
\author{R. Lauck}
\author{G. Siegle}
\affiliation{Max-Planck-Institut f{\"u}r Festk{\"o}rperforschung,
Heisenbergstrasse 1, D-70569 Stuttgart, Germany}

\author{A. Mu\~{n}oz}
\affiliation{MALTA Consolider Team, Departamento de F\'{\i}sica Fundamental II, and Instituto de Materiales y Nanotecnolog\'{\i}a, Universidad de La Laguna, La Laguna 38205, Tenerife, Spain}

\author{A.H. Romero}
\affiliation{CINVESTAV, Departamento de Materiales, Unidad
Quer$\acute{e}$taro, Quer$\acute{e}$taro, 76230, Mexico}

\author{A. Schindler}
\affiliation{NETZSCH-Ger\"atebau GmbH, Wittelsbacherstr. 42, D-95100 Selb, Germany}

\date{\today}

\begin{abstract}
We have measured the specific heat of zincblende ZnS for several isotopic compositions and over a broad temperature range (3 to 1100 K). We have compared these results with calculations based on \textit{ab initio} electronic band structures, performed using both LDA and GGA exchange-correlation functionals. We have compared the lattice dynamics obtained in this manner with experimental data and have calculated the one-phonon and two-phonon densities of states. We have also calculated mode Gr\"uneisen parameters at a number of high symmetry points of the Brillouin zone. The electronic part of our calculations has been used to investigate the effect of the 3$d$ core electrons of zinc on the spin-orbit splitting of the top valence bands. The effect of these core electrons on the band structure of the rock salt modification of ZnS is also discussed.
\end{abstract}

 \maketitle

%\draft

\email{M.Cardona@fkf.mpg.de}

\section{Introduction}\label{SecIntro}

This article is part of an on-going effort to investigate, both theoretically and experimentally, the specific heat of semiconductors and semimetals, with emphasis in the low temperature region where strong deviations from the Debye $T^3$ law take place.\cite{Gibin2005,Kremer2005,Serrano2006}  The reader will find more references to our recent work on the subject in Refs. \onlinecite{Romero2008} and \onlinecite{Cardona2009}.   Most of our past experimental and theoretical work has been performed for materials with different isotopic compositions, the exceptions being the mercury chalcogenides (HgX, X=S, Se, Te) for which only natural elements were considered. For the HgX materials, however, we made a  theoretical investigation of the effect of spin-orbit (\textit{s-o}) interaction on the specific heat and the phonon dispersion relations (this investigation was also carried out for Sb,\cite{Serrano2008},  Bi,\cite{Diaz2007},  and the lead chalcogenides\cite{Romero2008}).  The \textit{ab initio} calculations of $C_v$, based on the electronic structure as described by local density functionals, are able to account rather well for the experimental results.\cite{Heatpress} The availability of relativistic electronic band structures as a by-product of our calculations, suggested their use to investigate a number of interesting electronic and structural properties such as \textit{s-o} splittings (including the corresponding linear terms of the valence bands around \textbf{k}=0) , and also phase transitions under pressure.\cite{Cardona2009}

In this article we extend this work to ZnS, a material for which several isotopes of its constituent atoms and single crystals of its zincblende variety were available to us.\cite{Serrano2004}

ZnS is found in nature mostly in the zincblende (also called sphalerite, 3C) cubic structure (it actually gives its name to this structure-type) and constitutes the standard ore for the smelting of zinc. ZnS is also found as the mineral wurtzite, which lends its name to the canonical 2H structure in which several tetrahedral semiconductors crystallize (e. g. GaN, CdS, CdSe). As we shall see in the course of this work, the enthalpies of these two crystalline varieties of ZnS differ only very little, the 3C modification being slightly more stable than wurtzite  except at temperatures higher than $\sim$1200 K. \cite{Hu2008}  A transition from the 3C (zb) phase to the rocksalt (rs) structure takes place under pressure at about 15 GPa.\cite{Mujica2003}  The zb and wurtzite (wz) phases of ZnS differ only in the stacking order of crystal planes along one of the three-fold symmetry axes, a fact which probably accounts for the small difference in their enthalpies. A large number of similar phases (more than 200), with related stacking orders but larger primitive cells have been identified.\cite{Mardix1986}
They provide one of the most conspicuous examples of the phenomenon called polytypism\cite{Baronnet1992} We shall not be concerned in this article with this fascinating aspect of ZnS. From the practical point of view, we would also like to mention that polycrystalline ZnS is available commercially as windows and domes for infrared radiation detection devices.\cite{Harris1999} Because of its large electronic band gap, it is also useful for optoelectronic applications involving the short wavelength part of the optical spectrum\cite{Piquette1997, Bellotti1997}  although it has recently found a strong competitor in GaN.\cite{Nakamura2000}

We present in this paper measurements and \textit{ab initio} calculations of the specific heat $C_p$ of natural and isotopically modified zb-ZnS ( $^{64}$Zn, $^{68}$Zn, $^{32}$S, $^{34}$S) between 3 and $\sim$200 K (for ZnS with the natural isotope composition the work has been extended to 1100 K). We pay particular attention to the maximum in $C_p/T^3$  which occurs at about 22 K and obtain from the experimental data the logarithmic derivatives of $C_p$ vs. the isotopic masses of Zn and S.  These results are compared with the temperature dependence of $C_p$ in the manner followed in our earlier publications.\cite{Romero2008} Since the calculation of $C_v$ requires that of the phonon dispersion relations, we compare the results obtained for the latter from \textit{ab initio} electronic band structures with experimental results obtained by inelastic neutron scattering. We also present the calculated one-phonon density of states (DOS), explicitly separating the vibrational components of the two constituent atoms and provide the densities of two-phonon pairs with a total zero wavevector, including both sums and differences of two phonons. These DOS are relevant to the interpretation of Raman and ir spectra.  We compare the results obtained with extant second order Raman scattering data.

We have performed the \textit{ab initio} electronic calculations with and without  \textit{s-o} interaction. Because of the smallness of the \textit{s-o} effects, they barely influence the vibrational properties, contrary to what was found in materials containing heavier atoms.\cite{Romero2008,Cardona2009,Serrano2008,Diaz2007} It was found in Ref. \onlinecite{Cardona2009} that the \textit{s-o} splitting of core $d$ electrons of the cation provides a negative contribution to the \textit{s-o} splitting of the top valence bands (at the $\Gamma$ point) which, in the case of HgS, even reverses the sign of this splitting.\cite{Groves1963} We have investigated the possibility of such an effect in zb-ZnS and found that, while the \textit{s-o} splitting remains positive, the admixture of 3$d$ electrons of the Zn-core reduces its value. This reduction depends strongly on pressure. In the rs modification, however, parity prevents the 3$d$-zinc admixture with the 3$p$ wavefunctions of sulfur and such effects do not occur. In order to ascertain the consequences of this fact, calculations for rs-ZnS were performed at a pressure slightly above that of the phase transition. We noticed that the lack of $p$-$d$ admixture at the top of the valence band  is also responsible for converting the rs phase of ZnS into an indirect gap semiconductor (the zb gap is direct) in a way similar to that found earlier for AgBr and AgCl.\cite{Bassani1965} The lack of inversion symmetry in zb also generates \textit{s-o} splittings linear in \textbf{k} at the top of the valence bands. We evaluated them and compared them with previous semiempirical results.

Since some of the work just described requires calculations of the electronic structure as a function of pressure, we also evaluated the phonon dispersion relations of zb-ZnS at several different pressures below 15 GPa. We calculated the mode Gr\"uneisen  parameters at the $\Gamma$, X and L point of the Brillouin zone. From the parameters of the LO and TO modes at $\Gamma$ we evaluated the Gr\"uneisen parameter of the dynamical charge and found it to be negative, like for most other tetrahedral semiconductors (exception: SiC\cite{Anastassakis1998}), a fact that is usually interpreted as signaling the decrease  of ionicity with increasing pressure.

This article is structured as follows: In Section \ref{SecTheory}  we discuss the numerical procedures and codes employed for the \textit{ab initio} calculations. Section \ref{SecBandStruc} presents and discusses some details of the electronic band structure calculations that were used in the subsequent sections. Particular attention is paid to the effects of \textit{s-o} interaction mentioned above and also to the optimized lattice constants and bulk moduli which are obtained with the various codes employed. These results are compared with experimental data.

Section \ref{SecLattDyn} is devoted to the \textit{ab initio} calculations of the phonon dispersion relations and comparison with INS  and Raman data. For the purpose of calculating $C_v(T)$ we need the one-phonon  DOS which we present here together with its decomposition into Zn and S contributions. We also use it to estimate the contribution of isotopic disorder to the Raman linewidths for natural and for ZnS composed of  ($^{64}$Zn$_{0.5}\,^{68}$Zn$_{0.5}$)  ($^{30}$S$_{0.5}$\,$^{34}$S$_{0.5}$) isotopic mixtures. We also present the calculated two-phonon optical (\textbf{k}=0) DOS, for sums and for difference modes, and compare it with extant Raman data.

Although some information is already available in the literature, we evaluate in Sec.\ref{SecPressEffects}  the $T$=0 enthalpies of the three most important phases of ZnS vs. pressure. The results are presented as differences with respect to that of the most stable zb phase. At $T$ = 0 the enthalpy difference of the wz phase is basically independent of pressure (in recent work, this difference has been shown to decrease with increasing $T$  and to vanish for $T \sim$ 1200 K\cite{Hu2008}).
The zb to rs phase transition is calculated to take place for  $T$ $\sim$ 0  between 15.8 and 16.7 GPa, depending on whether the LDA or the GGA approximation is used for the exchange and correlation potential. As a by-product we report the pressure dependence of the phonon modes at $\Gamma$, L and X, and the corresponding Gr\"uneisen parameters. From those parameters for the LO and TO phonons at $\Gamma$, we derive the volume dependence of the dynamical effective charge and the corresponding Gr\"uneisen parameter.
Calculations of the electronic band structure vs. pressure reveal an anomalous decrease  of the \textit{s-o} interaction with increasing pressure, related to an increasing admixture of Zn-3$d$ core wavefunctions to those predominantly S-3$p$-like at the top of the valence bands.
Section \ref{SecLattDyn} presents the measurements of $C_p(T)$ for natural ZnS and for several isotopically modified samples. The results for natural ZnS are compared with various sets of rather incomplete data available in the literature and  with our \textit{ab initio} calculations.
The logarithmic derivatives of $C_p$ vs. isotopic mass are also compared with the derivative of $C_p$ vs. $T$  following the scheme used earlier for other binary semiconductors.\cite{Cardona2009,Serrano2006}
Section \ref{SecConclusions} presents the conclusions and summarizes the results.

\section{Theoretical Details}\label{SecTheory}

Semi-empirical calculations of the band structure and the lattice dynamics  of 3C-ZnS have been available for quite some time.\cite{Cohen1985,Serrano2004,Talwar1981} In the past few years, a number of articles reporting \textit{ab initio} studies of electronic, structural, and vibronic properties have appeared.\cite{Chen2008,Hu2008,Yu2009}  These articles cover some partial aspects of the work described here. Reference \onlinecite{Hu2008}, for instance, reports calculated values of $C_v$ only at five temperatures (300, 600, 900, 1200, and 1500 K) and thus misses the interesting region from 10 to 50 K in which the maximum of $C_v/T^3$ takes place. We thus decided to perform independent similar \textit{ab initio} calculations of all aspects related to the present work. The results will be compared with those of the Chinese researchers when appropriate.

The calculations were carried out using two different implementations of density functional theory\cite{Hohenberg1964,Kohn1965}, in order to ensure that our conclusions are independent of the basis wave expansion. We have also used different exchange-correlation functionals (LDA,GGA) for the same reason.

We have utilized a linear response approach\cite{Baroni1987,Gonze1997A,Gonze1997B}  together with an iterative minimization norm-conserving pseudopotential plane-wave method as implemented in the ABINIT package.\cite{Gonze2002} The local-density approximation (LDA) was employed for the exchange and correlation energy.\cite{Perdew1981} These pseudopotentials are single projector, ordinary norm conserving, based on the Troullier-Martins method.\cite{Hamann1979,Troullier1991,Payne1992} 12 and 6 "valence" electrons were used  for Zn and S, respectively. A 40 Ry cutoff was defined for the plane wave expansion and a 6$\times$6$\times$6 regular shifted \textbf{k}-point mesh. The calculation is well converged: increasing the plane-wave cutoff to 50 Ry and  the \textbf{k}-points mesh to 8$\times$8$\times$8 yields, on average, a change of  less then two percent in the phonon frequencies. For the phonon frequencies, we have used a \textbf{q}-mesh of 12$\times$12$\times$12, to guarantee a good coverage of the dispersion relations. The dynamical matrices were obtained from perturbation theory\cite{Gonze1997A,Gonze1997B} and a Fourier interpolation was used to increase the mesh sampling to improve the description of quantities such as vibrational density of states, heat capacity, etc..

In addition we have also performed calculations with the Vienna \textit{ab initio} simulation package (VASP), where density functional theory has been implemented by using the pseudopotential method to describe the electron wave functions (see Ref. \onlinecite{Kresse1996} and references therein). The exchange-correlation energy was either taken in the local density approximation (LDA) with the Ceperley-Alder prescription\cite{Perdew1981,Ceperley1980} or with the generalized gradient approximation, GGA, with the PBE prescription.\cite{Perdew1996} The projector-augmented wave (PAW) scheme\cite{Bloechl1994} was adopted and the Zn 3$d$ orbitals were included explicitly in the calculations. The set of plane waves extended up to a kinetic energy cutoff of 380 eV. This cutoff was found necessary to achieve highly converged results within the PAW scheme. We used a dense Monkhorst-Pack grid for the Brillouin zone (BZ) integration to ensure highly converged results (to about 1-2 meV per formula unit). We also employed an accurate algorithm during the calculations in order to assure very well converged forces in the calculation of the dynamical matrix. At each selected volume, the structures were fully relaxed to their equilibrium configuration through the calculation of the forces on atoms and the stress tensor.\cite{Manjon2006} In the relaxed equilibrium configuration, the forces are less than 0.002 eV/\AA\ and the deviation of the stress tensor from a diagonal hydrostatic form is less than 0.1 GPa.

In the case of VASP, highly converged  forces are required for the calculation of the dynamical matrix using the direct force constant approach (or supercell method).\cite{Parlinski} The construction of the dynamical matrix at the $\Gamma$ point of the BZ is particularly simple and involves separate calculations of the forces in which a fixed displacement from the equilibrium configuration of the atoms within the primitive unit cell is considered. Symmetry arguments were taken into account for reducing the number of such independent distortions. Diagonalization of the dynamical matrix provides both the frequencies of the normal modes and their polarization vectors. It allowed  us to identify the irreducible representation and the character of the phonons. Supercell calculations were carried out in order to obtain phonon dispersion curves.
We start with a 2$\times$2$\times$2 supercell for which the direct method delivers correct phonon frequencies  at special points of the BZ which are compatible within the selected supercell size. If the interaction range would cease within the selected supercell size the phonon frequencies would provide good interpolation values for all the wave vectors. (The LO/TO splitting cannot be included  using the direct method, but  the dynamical matrix can be supplemented with a non-analytical term, which depends on the Born effective charge tensors and the electronic dielectric constant, in order to obtain the LO/TO splitting.\cite{Pick1970}
Thermodynamic properties of the crystals at constant volume are determined by phonons.\cite{Parlinski} The ones measured at constant pressure require an anharmonic correction which has been included by hand.\cite{Heatpress}

We display in Table \ref{Table1} the lattice parameters we have used in our calculations which were found through energy minimization. The results presented below were obtained with either the ABINIT code (local density functional, LDA) or the VASP code (using either the LDA functional, or the generalized gradient approximation (GGA)).  Note that the experimental data for $a_0$ lie between the  GGA and the LDA results, the former being higher than the measured ones, the latter lower, as is usually the case.\cite{Chen2008,Nazzal1996}

\begin{table*}[h]
\caption{\label{Table1} Lattice parameters $a_0$ and first ($B_0$), second ($B'_0$) and third order ($B''$) bulk moduli at $p$ = $T$ = 0, obtained by energy minimization with the VASP code using GGA local density functional ( with and without \textit{s-o} interaction) and the LDA  (with \textit{s-o}) interaction. Calculated values were obtained through fits to the Birch-Murnaghan equation.\cite{Birch1947} The experimental values are taken from Ref.\onlinecite{Landolt}. results of our ABINIT-LDA values are also listed.}
\begin{ruledtabular}
\begin{tabular}{ccccc}
& $a_0$ (\AA) & $B_0$ (GPa) & $B'_0$ &	$B''_0$ (GPa$^{-1}$)	  \\ \hline
VASP-GGA, NOSO & 5.450 & 71.12 & 3.88 & 0.29$\times$10$^{-4}$\\
VASP-GGA, SO & 5.450 & 70.51 & 3.98 & -0.85$\times$10$^{-3}$\\
VASP-LDA, SO & 5.302 & 90.12 & 3.79 & -0.17$\times$10$^{-3}$\\
ABINIT-LDA, NOSO & 5.319 & 85.86 & 4.48 & -0.86$\times$10$^{-3}$\\
experiment & 5.404 & 76.9 & 4.4 & - \\
\end{tabular}
\end{ruledtabular}
\end{table*}

Correspondingly, the calculated bulk modulus $B_0$ is larger for the LDA than for the GGA calculations. The higher order bulk moduli $B'$ and $B''$ were obtained by fitting total energy calculations vs. the lattice parameter $a_0$ to the Birch-Murnaghan equation.\cite{Birch1947} $B'$ is found to have the typical values close to 4. $B''$ is rather small
and can be neglected for all practical purposes. Because of our previous interest in \textit{s-o} coupling effects on thermodynamic properties\cite{Diaz2007} we have actually performed GGA calculations with and without \textit{s-o}-interaction. This interaction does not alter the value of $a_0$  although it slightly increases $B_0$, i.e., the average acoustic phonon frequencies. This effect, however, is too small to be of interest.

\section{Electronic Band Structure}\label{SecBandStruc}

\begin{figure}[ht]
\includegraphics[width=8cm ]{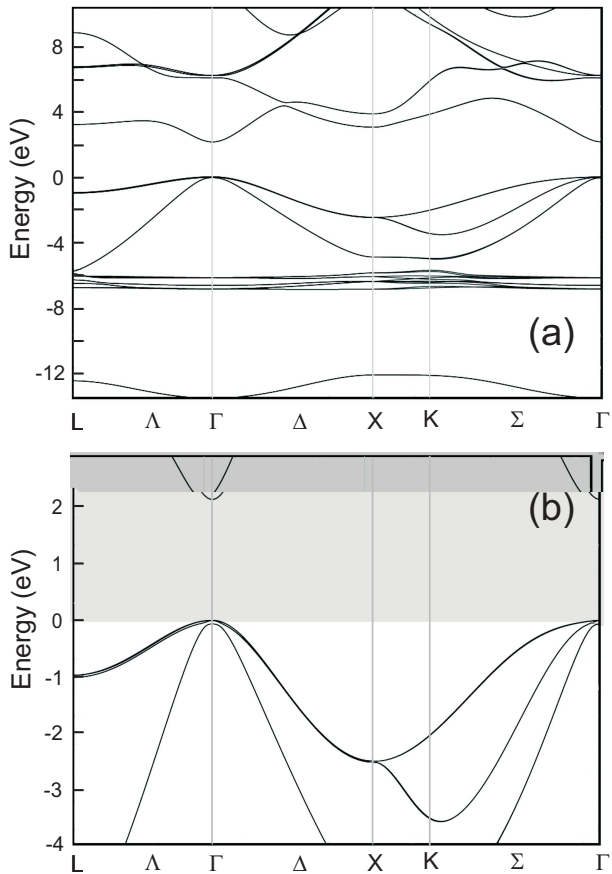}
\caption{(a) Electronic band structure of ZnS calculated with the VASP code using an LDA exchange correlation Hamiltonian. (a) includes the 3$d$ core-like electron bands of Zn.  (b) Magnified part of the band structure of ZnS which concentrates on the bands around the energy gap, so as to display the \textit{s-o} splittings.} \label{FigIII-1}
\end{figure}

Figure \ref{FigIII-1} displays the electronic band structure of zb-ZnS as calculated with the VASP code using the LDA exchange-correlation Hamiltonian. Figure \ref{FigIII-1}(a) shows the region around the direct energy gap at $\Gamma$ as well as the bands which correspond to the 3$d$ "semi-core" electrons of Zn (between 6 and 7 eV below the top of the valence band), which were explicitly included in the calculation. These bands are responsible for some of the interesting phenomena to be discussed below (the small \textit{s-o} splitting at the top of the valence bands and its anomalous dependence on pressure, the conversion of the direct into an indirect gap in the rs phase).

The calculated direct gap $E_0$ (2.19 eV, see Fig. \ref{FigIII-1}(b)) is considerably smaller than the one measured at 19K (3.85 eV,  cf. Ref. \onlinecite{Landolt}). This fact applies to all LDA and GGA calculations and is usually referred to as the "gap problem". It has been attributed to a discontinuity in the local density functional which takes place when crossing the gap.\cite{Godby1986} Recent calculations using the GW approximation yield for this discontinuity an energy of 2.05 eV.\cite{Shishkin2007,Tran2009} When adding this discontinuity to the 2.2 value of the gap (Fig. \ref{FigIII-1}(b)) one obtains  4.25 eV. This energy should be lowered by the electron phonon interaction  about 0.1 eV.\cite{Cardona2005RMP} The resulting value of 4.25 eV is somewhat higher than the experimental one. Smaller values of the discontinuity have been very recently suggested.\cite{Tran2009} The use of the experimental $a_0$ instead of the one obtained through energy optimization would also lower the calculated gap and thus decrease the remaining discrepancy.
We have also performed calculations using the GGA for the exchange-correlation functional, also including \textit{s-o} interaction. The gap obtained is 2.07 eV, somewhat smaller than the LDA gap. The difference (0.12 eV) is less than that estimated from the difference in lattice constants for the two functionals (see Table I) using the measured value for the pressure coefficient of $E_0$  ($dE_0 /dp$ = 0.057 eV/GPa, estimated LDA-GGA gap difference 0.37 eV).

\begin{figure}[ht]
\includegraphics[width=8cm ]{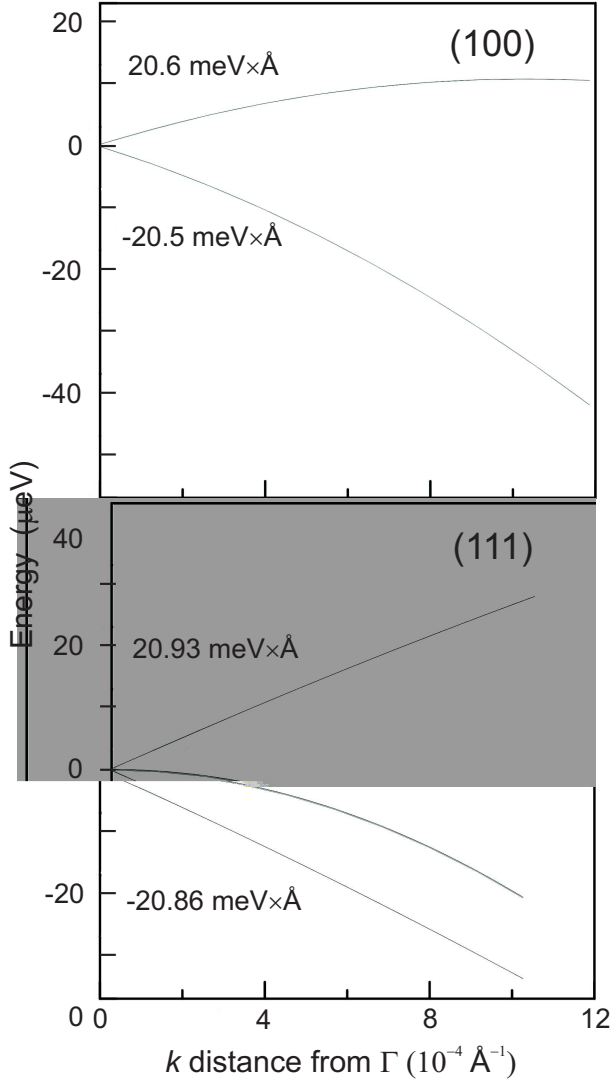}
\caption{Splitting of the $\Gamma_8$ valence bands for \textbf{k} along [100] and [111].  The fitted linear coefficient $C_k$ is given next to the corresponding curve. Their average is -20.7 meV$\times$\AA. Note that for \textbf{k} along [111] two of the quadruplet bands do not split. The splitting is $\pm C_k \cdot k$ for \textbf{k} along [100] and $\pm \sqrt{2} C_k \cdot k $ for \textbf{k} along [111].} \label{FigIII-2}
\end{figure}

We discuss next the value of the \textit{s-o} splitting $\Delta_0$  of the top valence bands at $\Gamma$. It was found in the calculations of Fig. \ref{FigIII-1} to be 0.062 eV, in agreement with the measured value ($\Delta_0$ = 0.067 eV, Ref. \onlinecite{Sobolev1978}) and with the \textit{ab initio} calculations of Carrier and Wei (0.066 eV).\cite{Carrier2004}   This value is rather small compared with estimates in the literature based on the splittings of the 3$p$ atomic levels of Zn and those of S (0.097 eV, Ref. \onlinecite{Cardona1968}). This fact suggests a significant contribution of the 3$d$
core electrons of Zn, similar to that which accounts for the sign reversal of $\Delta_0$ in HgS, CuCl, Zn0 and CuGaS$_2$ (Ref. \onlinecite{Cardona2009})  In order to verify this conjecture, and to obtain a quantitative estimate of the Zn-3$d$ contribution to $\Delta_0$ , we shall present in Sec. \ref{SecPressEffects} calculations of the electronic band structure of rs-ZnS. The crystal structure of rocksalt is centrosymmetric and thus no $p$-$d$ hybridization is possible at the $\Gamma$ point of the Brillouin zone. Consequently no compensation of the 3$p$ splitting at $\Gamma$ by the 3$d$ splitting of the Zn core levels should take place. We calculate for the rs phase a splitting of 0.112 eV, which corresponds to that listed in Ref. \onlinecite{Cardona1968}.

Another interesting effect of the spin-orbit interaction is the appearance of linear terms in \textbf{k} in the valence bands around $\Gamma$.\cite{Dresselhaus1955,Cardona1988,Cardona2009}  The existence of these terms follows from the lack of inversion symmetry through the coupling of the $\Gamma_8$ valence band states with the Zn-3$d$ "semi-core" states via the \textbf{k$\cdot$p} and the \textit{s-o} Hamiltonians. In Refs.  \onlinecite{Cardona1988} and \onlinecite{Cardona2009}, a semiempirical perturbation expression for the coefficient $C_k$  is given. For the zb II-VI compounds it reduces to:

\begin{equation}\label{Eq1}
C_k = -350 \frac{\Delta_{d,c}}{E(\Gamma_8^{\nu}) - E_{d,c}(\Gamma_{12})} \rm{(in\,\, meV\times\AA )}
\end{equation}

$\Delta_{d,c}$ is the \textit{s-o} splitting of the outmost cation $d$ levels and $E_{d,c}$ their energies, respectively.

Using for the energy denominator of Eq. (\ref{Eq1}) the value of 6.5 eV extracted from Fig. \ref{FigIII-1}   and the atomic Zn-3$d$ splitting of 0.4 eV (Table X in Ref. \onlinecite{Cardona1988}) we find $C_k$ = -21.5 meV$\times$\AA.\cite{Cardona1988conv}  This value of $C_k$ is slightly larger than that listed in Table X of Ref. \onlinecite{Cardona1988} for ZnS ($C_k$ = - 15.5 meV$\times$\AA) because a smaller value (6.5 eV) of the denominator of Eq. (\ref{Eq1}) has been used (9 eV was used in Ref. \onlinecite{Cardona1988}).

Our band structure calculations also enable us to obtain \textit{ab initio} the (linear) splittings under consideration. The corresponding results are shown in Figure \ref{FigIII-2} for \textbf{k} along [100] and [111]. Along [100] the $\Gamma_8$ bands split into two doublets. Along [111] $\Gamma_8$ splits into two singlets, and a doublet which is not affected by the \textit{s-o} interaction. From the initial slopes of the curves in Fig. \ref{FigIII-2}(b), an average $C_k$ = -20.7 meV$\times$\AA\ is obtained.

\section{Lattice Dynamics}\label{SecLattDyn}

\begin{figure}[ht]
\includegraphics[width=9cm ]{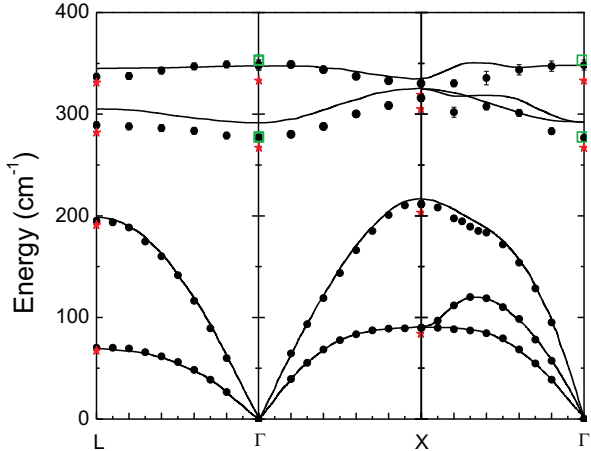}
\caption{(color online) Phonon dispersion relations of zb-ZnS, calculated with the ABINIT-LDA code, compared with the results of INS (dots, Ref. \onlinecite{Vagelatos1974}). The (green) squares represent the Raman frequencies measured at low temperatures. The stars display  \textit{ab initio} calculations  performed with the CASTEP-GGA code.\cite{Yu2009}} \label{FigIV-1}
\end{figure}

Figure \ref{FigIV-1} displays the phonon dispersion relations of rs-ZnS, calculated using the ABINIT code with the LDA approximation. We have also plotted in this figure the phonon frequencies measured at room temperature by Vagelatos \textit{et al.}   with INS.\cite{Vagelatos1974} Because of the anharmonic temperature shift  the measured frequencies are expected to be somewhat lower than calculated ones, a fact which seems to be supported by Fig. \ref{FigIV-1}.\cite{Cardona2005} This difference, however, may be enhanced by the fact that the LDA value of $a_0$  is smaller   than the experimental one (Table \ref{Table1}). We have also included in Fig. \ref{FigIV-1} frequencies calculated with the CASTEP code using the GGA.\cite{Yu2009}  These frequencies are also lower than those obtained with the LDA, even lower than those obtained experimentally,   a fact that suggests that the difference is mainly due to differences in the lattice parameters (larger for the GGA than either the experimental ones or those calculated with the LDA, Table \ref{Table1}). Note that the dispersion relations of the LO phonons are almost flat whereas those of the TO counterparts show a slight upward bending starting at the $\Gamma$ point. This bending, however, is much less than that found for  HgS  (Ref. \onlinecite{Cardona2009}), as corresponds to the smaller difference between the cation and anion masses.
The data of Fig. \ref{FigIV-1} at the $\Gamma$ point  allow us to calculate the transverse (or Born) effective charge $e^*$ using the standard expression:

\begin{equation}\label{Eq2}
\omega^2_{\rm{LO}} - \omega^2_{\rm{TO}} = \frac{4\,\pi\,(e^*\,e)^2}{M\,V_c\,\epsilon_{\infty}}
\end{equation}

where $\omega_{\rm{LO}}$  and $\omega_{\rm{TO}}$ are the corresponding phonon frequencies at the $\Gamma$ point (Raman frequencies), $M$ the reduced mass of the primitive cell (21.52 amu for natural ZnS), $V_c$  the volume of that cell, $\epsilon_{\infty}$ the electronic contribution to the long wavelength dielectric constant, and $e$ the elementary charge. Using the LDA values of the $V_c$  ($a_0^3$/4, as obtained from Table \ref{Table1}),   $\epsilon_{\infty}$ = 5.2 (Ref. \onlinecite{Landolt}, and the calculated Raman frequencies  we obtain $e^*$ = 1.76, which agrees reasonably with the values obtained from the experimental frequencies (2.01). In Sec. \ref{SecPressEffects} we shall discuss the dependence of this charge on pressure.

\begin{figure}[ht]
\includegraphics[width=8cm ]{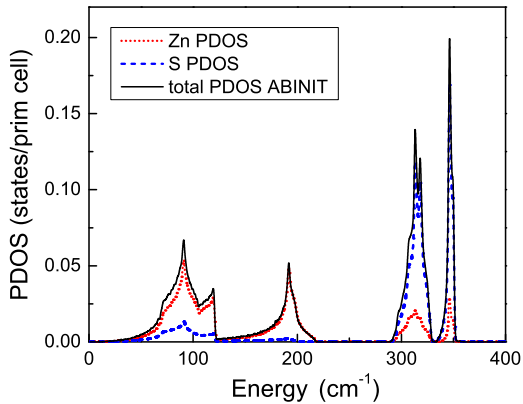}
\caption{(color online) Phonon DOS of ZnS obtained from the dispersion relations  shown in Fig. \ref{FigIV-1}. The DOS has been normalized so as to correspond to a total of 6 states per primitive cell. The partial components of the DOS (PDOS) corresponding to the Zn and the S vibrations are explicitly shown  (red) dotted line: Zn PDOS; (blue) dashed line: S PDOS). Notice the large gap between acoustic and optic phonons.} \label{FigIV-2}
\end{figure}

In order to calculate the temperature dependence of the specific heat we need the phonon DOS, which can be obtained by standard procedures from the dispersion relations of Fig. \ref{FigIV-1}. We display in Fig. \ref{FigIV-2} the total phonon DOS calculated for the ABINIT-LDA dispersion relations, together with the separate contributions of the Zn
and S vibrations. As expected, the acoustic branches are mainly Zn-like whereas the optic ones are mainly S-like.
These separate components are of interest for an estimate of the broadening induced by isotopic fluctuations in natural crystals or in those grown with strong isotopic disorder.\cite{Serrano2004,Cardona2005RMP,Cardona2009}  For this broadening to occur, there has to be significant phonon DOS at the frequency of interest, projected on the atom whose disorder is being considered. This effect is shown to be important for ZnSe with regards to the isotopic disorder of Zn.\cite{Goebel1999} In Ref. \onlinecite{Serrano2004} an unsuccessful attempt was made to identify such effect for the Raman phonons of ZnS. This can be understood on the basis of the disperson relations of Fig. \ref{FigIV-1} and the DOS of Fig. \ref{FigIV-2}. At the frequency of the TO phonons at $\Gamma$ the DOS  vanishes. This is not so easy to conclude for the nearly flat LO branch. However, the experimental results of Serrano \textit{et al.} (their Table III) suggest that isotopic disorder is absent for the LO ($\gamma$) phonons of ZnS.\cite{Serrano2004}

Phonon DOS have been obtained previously with semiempirical techniques.\cite{Talwar1981,Vagelatos1974} They agree semiquantitatively with the \textit{ab initio} results reported here and with the recent \textit{ab initio} work of Yu \textit{et al.}.\cite{Yu2009} These authors, however, did not report the partial components of the DOS.

\begin{figure}[ht]
\includegraphics[width=7.5cm ]{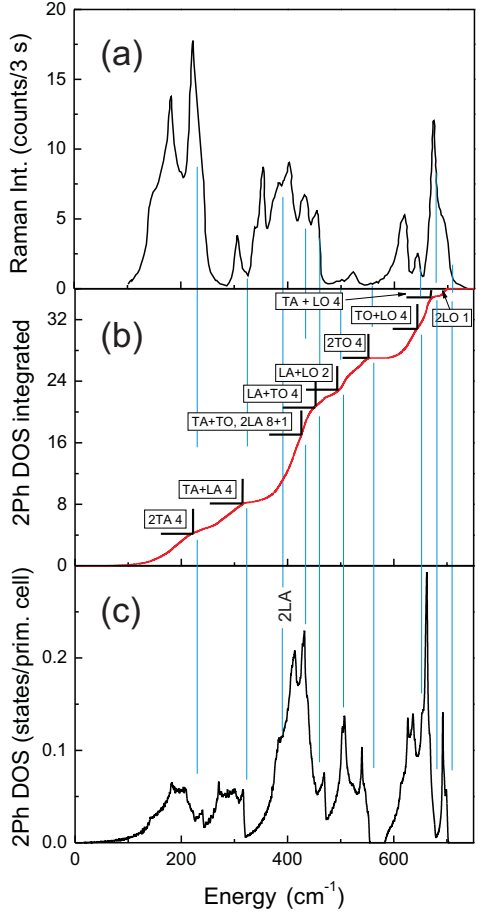}
\caption{(color online) The lower vignette (c) displays   the density of states which corresponds to the sum of two phonons with equal but opposite wavevector (2Ph DOS). The middle vignette (b) represents the integral of that 2Ph DOS (see text) whereas the upper vignette (a) shows the corresponding 2Ph Raman spectrum.\cite{Serrano2004} The 2Ph DOS is normalized to 36 states per primitive cell.} \label{FigIV-3}
\end{figure}

While several calculations of the one-phonon DOS are already available, the spectral distribution of two-phonon (2Ph) DOS (with   \textbf{k}$_1$ + \textbf{k}$_2$ = 0, as imposed by optical absorption and Raman scattering) is not that readily available. Semi-empirical calculations of the 2-phonon DOS are presented in Ref. \onlinecite{Talwar1981} in the form of histograms which have poor resolution. Semiempirical two-phonon DOS spectra were used in Ref. \onlinecite{Serrano2004} to interpret the changes in Raman phonon line widths observed upon application of pressure and to assign structure in the two-phonon Raman spectra. We present here, for the sake of completeness, such calculations performed for the \textit{ab initio} band structure of Fig. \ref{FigIII-1}. We display in Fig. \ref{FigIV-3}(a) the DOS that corresponds to the sums of two phonons (with total \textbf{k}=0), normalized to 36 states per primitive cell. In order to identify the origin of the various bands involved, we also plot the integral of that DOS vs. frequency from  $\omega_1$ + $\omega_2$  equal to zero to its maximum: each plateau in the integrated DOS signals the end of a 2Ph band. These bands, and their weights, are labeled  in the integrated curve. As an example we mention that the lowest band (from 0 to $\sim$ 220 cm$^{-1}$) corresponds to two TA phonons, whereas the highest, rather narrow band corresponds to two LO phonons. We have plotted at the top of Fig. \ref{FigIV-3} the Raman spectra measured in this 2Ph region.\cite{Serrano2004}  The various bands observed can, in this manner, be assigned to 2Ph combinations. The vertical (blue) lines  try to establish correspondences between structure in the Raman spectra and in  the 2Ph DOS. A more detailed assignment to van Hove singularities (critical points) has been attempted in Ref.\onlinecite{Serrano2004}.

\begin{figure}[ht]
\includegraphics[width=7.5cm ]{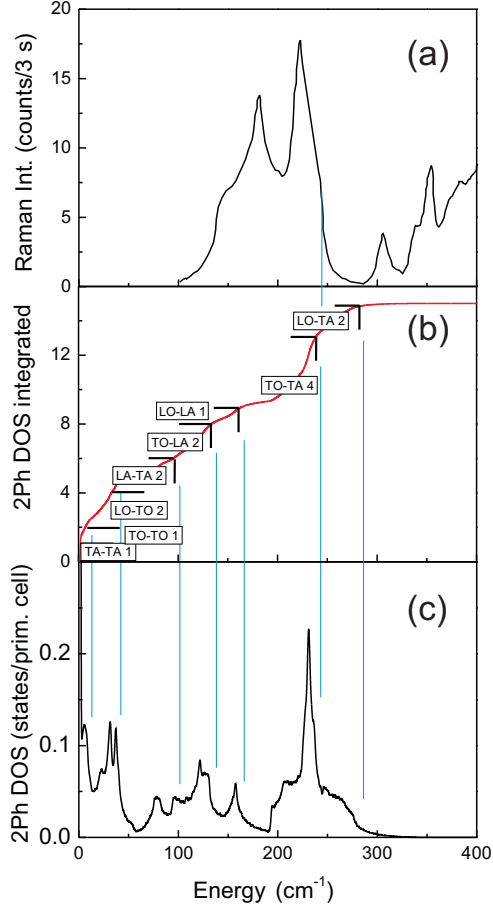}
\caption{
Similar to Fig. \ref{FigIV-3} but for the DOS which corresponds to 2Ph differences. The normalization is to 15 phonon difference states per primitive cell.} \label{FigIV-4}
\end{figure}

Similar plots are shown in Fig. \ref{FigIV-4} for the calculated 2Ph difference DOS.  The corresponding Raman spectra vanish at low temperatures because of the appropriate Bose-Einstein statistical factor (or, in other words, because no phonons are available for being annihilated). The spectrum on the top of Fig. \ref{FigIV-3}(b) was obtained at 350 K, a temperature at which annihilation of TA phonons becomes possible. In this manner, the strong peaks at $\sim$ 220 cm$^{-1}$ was identified as corresponding to the creation of a TO and the annihilation of a TA phonon. In spite of the additional structure in the 2Ph difference DOS shown in Fig. \ref{FigIV-4}, no structure in the Raman spectra has been unambiguously identified as belonging to difference modes. We believe that the DOS of Fig. \ref{FigIV-4}, combined with detailed measurements of the temperature dependence of the Raman spectra, should help in identifying additional 2Ph difference structure.

\section{Specific Heat}\label{SecHeatCap}

Several publications have already presented partial aspects of the specific heat of zb-ZnS. Low temperature data  were early reported by Clusius and Harteck (19 K $<T<$ 196 K) and by Martin (4 K $<T<$ 20 K).\cite{Clusius1928,Martin1955} More recent data were presented by Birch  but only for $T <$ 12 K.\cite{Birch1975} Most of the existent high temperature data (above room temperature up to 1200 K) were obtained for polycrystalline
material of the type used in ir-windows (Cleartran$^{\circledR}$ multispectral zinc sulfide).\cite{Harris2008} Here we present experimental results for $C_p(T)$ of four zb-ZnS samples grown by vapor phase transport (for details of the growth see Ref. \onlinecite{Serrano2004}) with different isotopic compositions: $^{\rm nat}$Zn$^{\rm nat}$S (i.e. Zn and S with natural isotope composition), $^{68}$Zn$^{32}$S, $^{64}$Zn$^{32}$S, $^{68}$Zn$^{34}$S$_{0.5}$ $^{32}$S$_{0.5}$). The $^{68}$Zn$^{34}$S$_{0.5}$ $^{32}$S$_{0.5}$  sample should behave like  $^{68}$Zn$^{33}$S and thus, for simplicity, it was assumed to have this composition. The measurements covered the temperature range from 3 to 1100K and so did also our corresponding \textit{ab initio} calculations.

The specific heat $C_p$   between 2 and 280 K was measured with a PPMS system\cite{PPMS} as described before.\cite{Cardona2005,Gibin2005,Kremer2005}
Between room temperature and  1100 K the heat capacity was determined with a
DSC 404 F1 Pegasus differential scanning calorimeter.\cite{Netzsch}

\begin{figure}[ht]
\includegraphics[width=8cm ]{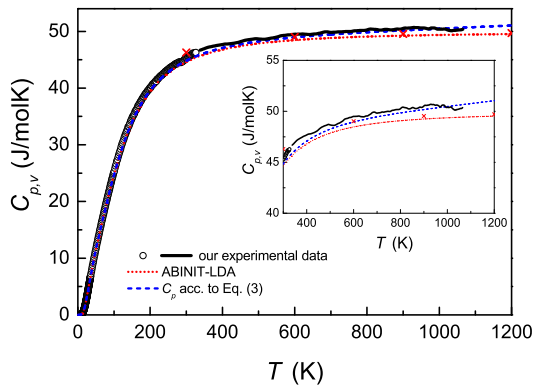}
\caption{(color online) Measured ($\circ$ and solid line) and calculated values of the specific heat of zb-ZnS in the 3-1200 K temperature range ((red) dotted line: ABINIT-LDA $C_v$; (blue) dashed line: $C_p$ calculated according to Eq. (\ref{Eq3}). The four points represented by (red) {\large{$\times$}} correspond to the calculations of Hu \textit{et al.}.\cite{Hu2008} The inset displays our high temperature data  in an enlarged scale.}\label{FigV-1}
\end{figure}

We show in Fig. \ref{FigV-1} the $C_p(T)$ data obtained for a sample with natural isotopic composition  in our whole $T$ range, together with our ABINIT-LDA calculations of $C_v(T)$. The difference that appears above $\sim$600 K, approximately linear in $T$, can be attributed to the thermal expansion and has been estimated with the standard expression:\cite{Kremer2005}

\begin{equation}\label{Eq3}
C_p(T) - C_v(T) = a_v^2(T)\,B\,V_{mol}\,T
\end{equation}

where $a_v(T)$ is the  thermal volume-expansion coefficient taken from Ref. \onlinecite{Roberts1981}, which
in the temperature region under consideration increases slightly linear with the temperature.
$B$ is the isothermal bulk modulus and $V_{mol}$ the molar volume.
$C_p$ calculated from our $C_v$  using Eq. (\ref{Eq3}) is also shown in Fig. \ref{FigV-1}. It agrees rather well with the experimental results. For comparison, we have included the $C_v$  recently calculated \textit{ab initio}  by Hu \textit{et al.}  using the CASTEP code with the GGA approximation for the exchange and correlation potential at four temperatures.\cite{Hu2008} These points agree rather well with our $C_v$ calculations. A small discrepancy appears at 300 K which could be due to the use of GGA vs. LDA: GGA leads to a larger lattice constant (Table \ref{Table1}) and thus to lower phonon frequencies which correspond to higher values of $C_v$.

\begin{figure}[ht]
\includegraphics[width=8cm ]{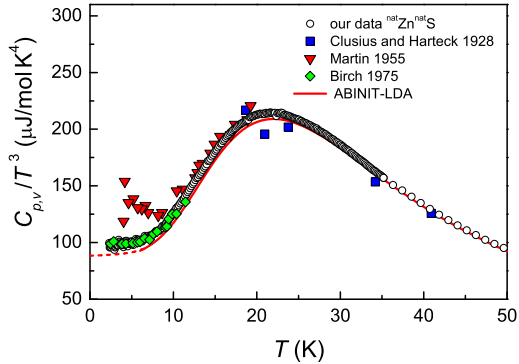}
\caption{(color online) Temperature dependence of $C_p/T^3$  in the $T$ region from 3 to 50 K measured on a ZnS sample with natural isotope abundance. Low temperature data available in the literature are also plotted.  The results of our \textit{ab initio} ABINIT-LDA calculations are represented by the (red) solid line; extrapolation to $T \rightarrow$ 0 K by the (red) dashed line.} \label{FigV-2}
\end{figure}

The isotopic composition is not expected to significantly affect the high temperature results of Fig. \ref{FigV-1}, which must tend to the Petit and Dulong  value of 49.9 J/mol\,K.\cite{Petit1819} The isotope effect should be seen best at the maximum in $C_p/T^3$ which, for zb-ZnS is expected to occur (and is in fact observed in the experimental data, cf. Fig. \ref{FigV-2})  at around 21.5 K. This peak occurs in zb-type materials at a temperature that corresponds to $\sim$1/6 of the frequency of the TA band in the DOS, expressed in temperature units, 130 K according to Fig. \ref{FigIV-1}.

\begin{figure}[ht]
\includegraphics[width=8cm ]{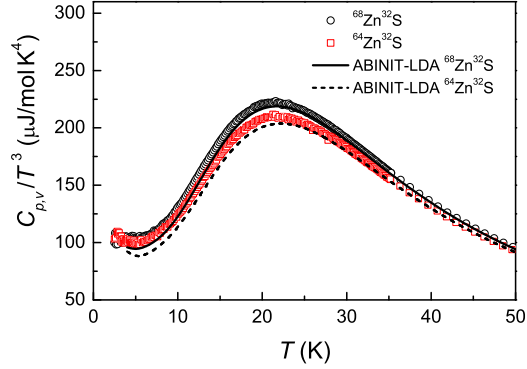}
\caption{(color online) Temperature dependence of $C_p/T^3$  in the $T$ region from 3 to 50 K measured for two zb-ZnS samples with different isotopic compositions as indicated. Also given are the results of our \textit{ab initio} ABINIT-LDA calculations.} \label{FigV-4}
\end{figure}

The measured effect of the substitution of  $^{68}$Zn for $^{64}$Zn on the maximum of $C_p/T^3$ vs. $T$ is shown in Fig. \ref{FigV-4}.  The $C_p/T^3$ maximum is too shallow to be able to investigate the dependence of its temperature on isotopic mass, although it seems to shift up in temperature with decreasing mass, as expected. The value of $C_p/T^3$  at its maximum, however, is shown to increase with increasing mass. This increase is nicely reproduced by our ABINIT- LDA calculations also shown in Fig. \ref{FigV-4}.

\begin{figure}[ht]
\includegraphics[width=7.5cm ]{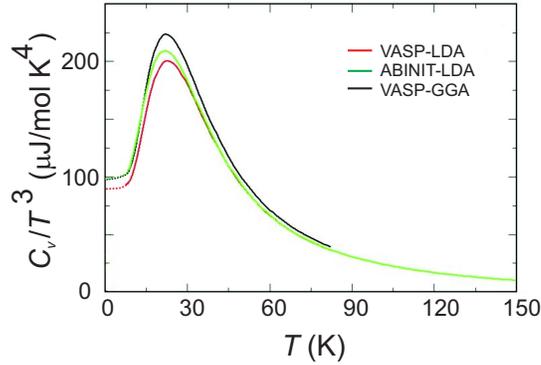}
\caption{(color online) $C_v/T^3$ for $^{\rm nat}$Zn$^{\rm nat}$S as calculated by ABINIT-LDA, and VASP-GGA and VASP-LDA, as indicated in the inset.} \label{FigV-3}
\end{figure}

Before proceeding with the isotopically modified samples we would like to recall that although most of the calculations presented here have been performed with the ABINIT-LDA code, we have also used the VASP code with either GGA or LDA exchange-correlation potentials. In order to compare the $C_v(T)$ results obtained with the three codes we display in Fig. \ref{FigV-3} the corresponding computations for zb-ZnS with the natural isotopic abundances.
The difference between the two LDA calculations ( $\sim$3\%) is not to be taken seriously, in view of the differences inherent to the VASP and ABINIT programs.

More interesting is the higher value of the maximum in $C_v/T^3$ obtained from calculations using the GGA Hamiltonian. Since such calculations yield a larger lattice parameter than those based on LDA (see Table I), one can conjecture that the higher value of $C_v/T^3$ is related to the average lower phonon frequencies that result from the larger GGA lattice parameters. The measured values of $C_v/T^3$ at the low temperature maximum fall between the LDA and the GGA calculations, as expected from the corresponding lattice parameters.
The dotted values displayed in Fig. \ref{FigV-3}  below $\sim$8 K have been drawn by free-hand: in this region the accuracy of our calculations does not allow reliable results.

In Figure \ref{FigV-5} we display the logarithmic derivative of $C_p/T^3$ vs. the mass of the Zn isotope $^{\rm M}$Zn evaluated for the experimental as well as the theoretical data of Fig. \ref{FigV-4}. Because of computational limitations, the calculated derivatives are meaningful only down to $\sim$ 10 K. The dashed continuation of this curve to the value 1 has been performed by free hand. The value 1 is obtained using the expressions (derived from Debye's theory):

\begin{figure}[ht]
\includegraphics[width=8cm ]{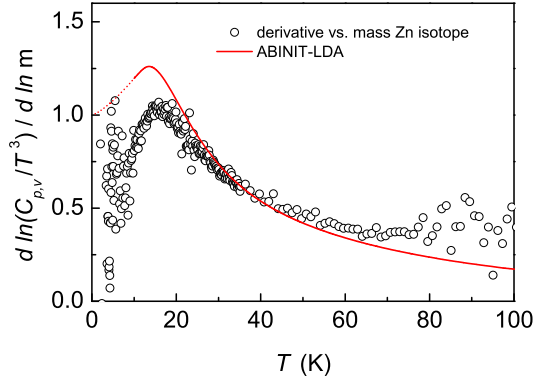}
\caption{(color online) Logarithmic derivative of the specific heat versus Zn mass as measured and \textit{ab initio} calculated by ABINIT-LDA for zb-ZnS. The dashed part of the calculated curve has been extrapolated free-hand so as to agree with the value of Eq.(\ref{Eq4}) at $T$ = 0.} \label{FigV-5}
\end{figure}

\begin{equation}
    \label{Eq4}
    \frac{d\ln(C_{v}/T^3)}{d\ln M_{\rm Zn}} = \frac{3}{2}\,\frac{M_{\rm Zn}}{M_{\rm Zn}+M_{\rm S}} \approx 1.0\\
\end{equation}

\begin{equation}
    \frac{d\ln(C_{v}/T^3)}{d\ln M_{\rm S }} = \frac{3}{2}\,\frac{M_{\rm S }}{M_{\rm Zn}+M_{\rm S}} \approx 0.5
    \label{Eq5}
\end{equation}

The maximum of the measurements of Fig. \ref{FigV-5} takes place at 17 K. In most of our earlier work on this topic we have evaluated the ratio of the maximum frequency in the corresponding branch of the Ph-DOS to the temperature of the maximum in $C_v/T^3$. In the case of Zn. the corresponding maximum is that of the TA phonons, which corresponds to 130 K. The ratio (130 K/17 K) = 7.6 is rather close to that found for ZnO (7.3, Ref. \onlinecite{Serrano2006}), a value which is also close to that calculated on the basis of a single frequency Einstein model.\cite{Cardona2007}

\begin{figure}[ht]
\includegraphics[width=8cm ]{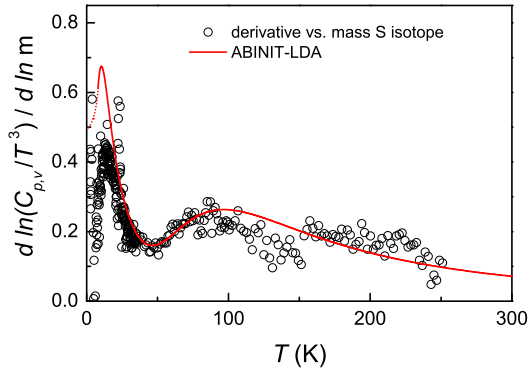}
\caption{(color online) Logarithmic derivative of the specific heat versus S mass as measured and \textit{ab initio} calculated by ABINIT-LDA for zb-ZnS. The dashed part of the calculated curve has been extrapolated free-hand so as to agree with the value of Eq.(\ref{Eq5}) at $T$ = 0.} \label{FigV-6}
\end{figure}

Figure \ref{FigV-6} displays the logarithmic derivative of $C/T^3$ vs. the mass of the sulfur isotope, evaluated from experimental as well as theoretical data. Again, the theoretical curve has been extrapolated free-hand in the low temperature region so as to agree at $T$ = 0 with Eq. (\ref{Eq5}). The calculations, as well as the experimental results, show two maxima which can be represented by two Einstein oscillators. The lowest maximum corresponds to TA phonons and its appearance reflects the existence of a considerable Zn component in the TA vibrations. The maximum at about 90 K can be assigned to the optical phonons (average frequency $\sim$ 330cm$^{-1}\sim$ 478 K). The ratio 478 K / 90 K = 5.3 is also close to that found for the corresponding maximum in ZnO (5.1)\cite{Serrano2006}).

\begin{figure}[ht]
\includegraphics[width=8cm ]{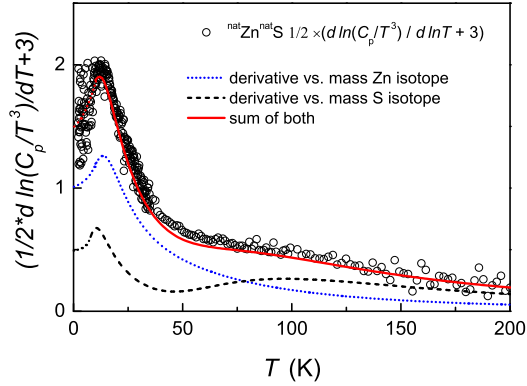}
\caption{(color online) Relationship between the derivative of $C_v/T^3$ vs. temperature and the corresponding derivatives with respect to the isotopic masses of Zn and S. The dotted lines were calculated from the heat capacities by varying the isotope mass of either Zn or S using the ABINIT-LDA code. The points are derived from the measured $C_p/T^3$ vs. $T$ by taking numerical  derivatives with respect to the isotope mass or with respect to temperature.} \label{FigV-7}
\end{figure}

In previous work on monatomic crystals we found a connection between the logarithmic derivative of  $C_v/T^3$ versus $T$ and the corresponding derivative vs. isotopic mass.\cite{Gibin2005}  For binary materials, a similar connection was shown to hold provided one adds the two derivatives with respect to each of the isotope masses.\cite{Cardona2007}
Figure \ref{FigV-7} shows that this connection is also valid for ZnS. The experimental points in the figure were obtained using the calculated mass derivatives and the calculated as well as the measured derivatives vs. $T$.

\section{Pressure Effects on the Structure and Lattice Dynamics}\label{SecPressEffects}

The availability of reliable \textit{ab initio} electronic structures enables us to tackle a few outstanding questions concerning the total energy
and the lattice dynamics as a function of pressure. The first such question appeared in Sect. III in connection with the spin - orbit splitting at the top of the valence band in zb-ZnS. This splitting was calculated \textit{ab initio}
to be $\Delta_0$ = 0.062 eV, in reasonable agreement with the measured value (0.067 eV., Ref. \onlinecite{Cardona1968}). It was pointed out that this splitting was considerably smaller than that predicted on the basis of the 2$p$ wave functions of S and the 3$p$ of Zn. The reduction of \textit{s-o} splitting was conjectured to be due to the admixture of 3$d$ core-like wave functions of Zn (this admixture is also responsible for the linear \textbf{k}-terms of Eq. (\ref{Eq1})). In order to test this conjecture we calculated the dependence of $\Delta_0$ on volume, which we express as:

\begin{equation}\label{Eq6}
\gamma_{\Delta} = -\frac{d ln \Delta_0}{d ln V}.
\end{equation}

For most tetrahedral semiconductors $\gamma_{\Delta}$ is positive and lies around 0.5.\cite{Cerdeira1970}
This fact can be qualitatively understood as resulting from the wave function renormalization that takes place when the volume of the unit cell is changed. A uniform renormalization would lead to $\gamma_{\Delta}$ =+1. The resistance of the core to the penetration of the wavefunction reduces this value to the typical values around +0.5 just mentioned. For ZnS, however, a change in volume is expected to alter the 3$d$-Zn admixture: a decrease in the volume should increase the 3$d$ admixture at the top of the valence band. Because of the negative contribution of 3$d$ electrons to $\Delta_0$ this effect should result in a decrease in $\gamma_{\Delta}$, and possibly even a sign reversal. Our \textit{ab initio} VASP calculations yielded indeed $\gamma_{\Delta}$ = -0.58. The value $\gamma_{\Delta}$ = -0.12 had already been obtained by Cerdeira \textit{et al.} \cite{Cerdeira1970} from non-selfconsistent (not truly \textit{ab initio}) Korringer-Kohn-Rostocker (KKR) calculations.

The $p$-$d$ admixture responsible for the anomalies just mentioned should not take place in crystals with inversion symmetry, e.g. in the rock salt phase of ZnS (rs-ZnS). We thus performed band structure calculations for rs-ZnS and obtained for $p$=0 (the rs-ZnS phase is not stable at this pressure but this should not affect our argument) $\gamma_{\Delta}$ = +0.60 thus confirming our conjecture concerning the nature of $\Delta_0$ in both phases of ZnS. As already mentioned in Sect. \ref{SecBandStruc}, the value of  $\Delta_0$  calculated for the rs phase is 0.112 eV, much larger than that of the zb phase.

In order to proceed with our studies of the effects of pressure on the electronic and vibrational properties of ZnS we have investigated the region of stability of the zb and also the wurtzite  (wz) phase concerning the transition to the rs phase.\cite{cinnabar}

\begin{figure}[ht]
\includegraphics[width=7.5cm ]{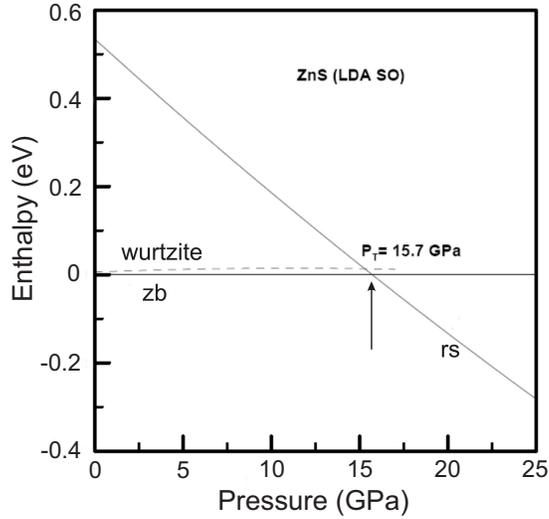}
\caption{Enthalpy difference between the rs and wz phases of ZnS and that of the zb phase (the latter is assumed to have zero enthalpy) at T= 0 K (neglecting zero-point vibrations), as calculated with the VASP-LDA code.} \label{FigVI-1}
\end{figure}

A number of calculations of the pressure at which the zb to rs transition and also the wz to rs transition take place have already appeared in the literature. Most of them use the LDA Hamiltonian\cite{Qteish1998,Martinez2006,Duranduru2009,Ves1990,Jaffe1993}, one of them uses the GGA  Hamiltonian.\cite{Hu2008} The extant LDA calculations indicate  transition pressures from zb-ZnS to rs-ZnS between 13 and 19.5 GPa, whereas the GGA calculation  gives 17.2 GPA.\cite{Hu2008} Experimental values lie between 14.7 and 18.1 GPa.\cite{Qteish1998}  Because of the broad range of transition pressures which have been reported, we decided to perform calculations using both, LDA and GGA Hamiltonians. The obtained enthalpy differences between rs,  wz and the zb phase are shown in Fig. \ref{FigVI-1} for the VASP-LDA calculation (including \textit{s-o} interaction). In this figure the zero temperature enthalpy of the wz modification lies about 0.008 eV per primitive cell above that of zb (it decreases slightly towards zero pressure) The zb to rs transition is predicted to occur at 15.7 GPa, a pressure that falls in the middle of the thus far reported range (13 to 19.5 GPa). Our corresponding VASP-GGA calculation yielded a wz phase also 0.008 eV per primitive cell above zb and a
transition pressure of 16.7 GPa. One can also here conjecture that the higher transition pressure of the GGA calculation is related to the larger lattice parameter.

\begin{figure}[ht]
\includegraphics[width=7.5cm ]{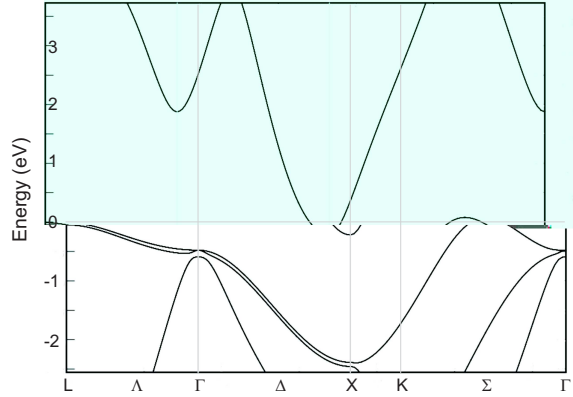}
\caption{Band structure of rs-ZnS calculated for $p$=16 GPa using the ABINIT-LDA code (\textit{s-o} coupling included). Notice the overlap between conduction and valence bands which, as explained in the text, is spurious and disappears once the appropriate "scissors operator" is applied.} \label{FigVI-2}
\end{figure}

In order to investigate additional effects of the inversion symmetry present in the rs structure we calculated the electronic band structure of rs-ZnS at a pressure $p$ = 16.0 GPa,
at which this phase is stable (see Fig. \ref{FigVI-1}). The results are displayed in Fig. \ref{FigVI-2}. The \textit{s-o} splitting of the valence bands at the $\Gamma$point is found to be 0.12 eV, considerably higher than that of the zb modification, as already discussed. The most conspicuous difference between the band structure of Fig. \ref{FigVI-2} and that of Fig. \ref{FigIII-1} is the lack of a gap: The minimum of the conduction band is found at the X-point, at an energy of -0.2 eV. The maximum of the valence bands is along the $\Gamma$-K direction, at an energy of 0.1 eV. Thus there seems to be an overlap of about 0.3 eV between valence and conduction bands. At this point, however, we must keep in mind the "gap problem" already mentioned in Sect. \ref{SecBandStruc}. For the case of the zb-ZnS modification, the calculated valence and conduction bands had to be pulled apart by about 2.05 eV (Our Refs. \onlinecite{Tran2009} and \onlinecite{Cardona2005}) in order to correct this problem. Using the same correction for the band structure of Fig. \ref{FigVI-2} as a so-called "scissors operator", a gap of $\sim$1.75 eV opens and rs-ZnS becomes an indirect gap semiconductor, with the valence band maximum along $\Gamma$-K ($\Sigma$ direction) and the conduction band minimum at X.
A band structure calculation for rs-ZnS has already appeared in the literature.\cite{Ves1990} It was performed with the linear muffin tin orbital method (LMTO) also using an LDA Hamiltonian and was thus also affected by the gap problem.
The authors of Ref. \onlinecite{Ves1990} performed transmission measurements for a rs-ZnS crystal obtained under pressure. They found an indirect gap of $\sim$1.9 eV, in good agreement with the one we obtained after applying the "scissors operator" as mentioned above

It is easy to explain in a qualitative manner the remarkable difference between the band structures of Figs. \ref{FigVI-2} and \ref{FigIII-1} as due to the effect of inversion symmetry (which is lacking for zb). Inversion symmetry prevents the hybridization of the $p$-like top of the valence bands with the 3$d$ core states at the $\Gamma$ point of zinc since both sets of states have opposite parity. When moving away from the $\Gamma$ point, either towards L or K, parity is no longer a good quantum number, the $p$ and $d$ states mix and the $p$-like top of the valence band is pushed up, as shown in Fig. \ref{FigVI-2}. This suffices to make the gap indirect.

The effect just described is similar to that encountered in rs-structure AgCl and AgBr (Ref. \onlinecite{Bassani1965,Scoop1965}) which also have a valence band maximum along $\Sigma$  and an indirect gap around 3 eV. The minimum of the conduction band of these materials, however, is at the $\Gamma$point at $p$=0. Application of pressure (our calculations were performed for $p$=16 GPa) would probably raise the $\Gamma$
minimum and lower that at X (Ref. \onlinecite{Paul1961})   bringing the band structure of AgCl and AgBr even more in line
with that  of  Fig. \ref{FigVI-2}.

\begin{figure}[ht]
\includegraphics[width=7.5cm ]{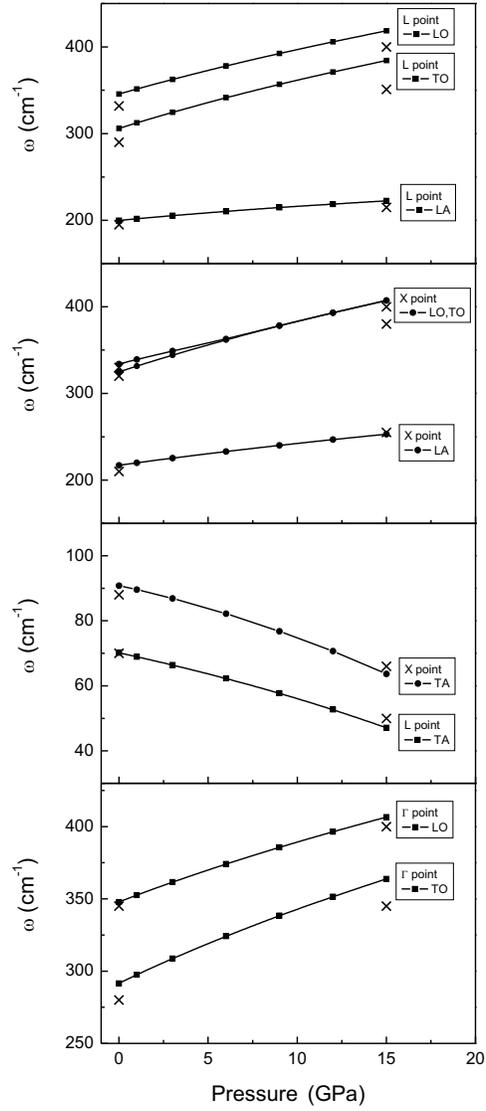}
\caption{(solid lines) Pressure dependence of the phonons of natural ZnS as calculted with the \textit{ab initio} ABINIT-LDA code at the indicated high symmetry points in the Brillouin zone.  Experimental data taken from Ref. \onlinecite{Weinstein1984}. are represented by  $\times$} \label{FigVI-3}
\end{figure}

We conclude this section by showing calculations of the pressure dependence of the phonons of natural ZnS calculated with the \textit{ab initio} ABINIT-LDA code at several high symmetry points ($\Gamma$, X and L). The figure displays the calculated points (full circles ) and also experimental points obtained by Weinstein.\cite{Weinstein1977} The solid lines are quadratic fits to the calculated results.  Of particular interest is the supralinear decrease of the TA frequencies with pressure found at the edge of the Brillouin zone (X and L points). This decrease seems to be qualitatively related to the rs phase transition\cite{Weinstein1977} The agreement between the experimental and calculated results of Fig. \ref{FigVI-3} is remarkably good.

From the coefficient of the linear terms of the fits in Fig. \ref{FigVI-3} we have calculated the corresponding mode Gr\"uneisen parameters  using the calculated bulk modulus $B_0$ = 85.86.   They are listed in Table \ref{Table2}  together with extant experimental data. Again, the agreement is good. Notice that at the $\gamma$ point  $\gamma_{\rm TO}$  is larger than
$\gamma_{\rm LO}$, a fact common to most zb-zype semiconductors.\cite{Anastassakis1998} These Gr\"uneisen parameters can be used to calculate the dependence of the transverse effective charge on volume, and the corresponding Gr\"uneisen parameter $\gamma_e$, using the expression obtained from Eq. (\ref{Eq2}):

\begin{equation}\label{Eq7}
\gamma_e = - \frac{d\,ln\,\epsilon^*}{d\,ln\,V} = 0.5 - \frac{d\,ln\,\epsilon_{\infty}}{2\,d\,ln\,V} + \frac{ \gamma_{\rm{LO}}\omega_{\rm {LO}}^2 -\gamma_{\rm{TO}}\omega_{\rm {TO}}^2}{\omega_{\rm {LO}}^2-\omega_{\rm {TO}}^2}.
\end{equation}

The logarithmic derivative of $\epsilon_{\infty}$  for ZnS has been reported by  Vedam and Schmidt  to be +0.36.\cite{Vedam1966} Using this value and those from Table \ref{Table2} into Eq. (\ref{Eq7}) we find $\gamma_e$ = -0.84, a result rather similar to that found for other zb-type semiconductors (exception: SiC).\cite{Anastassakis1998} The negative value of $\gamma_e$ implies an increase in ionicity with increasing volume. Using the experimental values of the $\gamma$ and $\omega$ in Table \ref{Table2} we obtain $\gamma_e$ = -0.44.

\begin{table*}[h]
\caption{\label{Table2} Mode Gr\"uneisen parameters at the $\Gamma$, L and X point as obtained from the pressure dependence of the phonon energies obtained by ABINIT-LDA calculations compared with available experimental data. If we use the experimental value $B_0$= 76.9 instead of the ABINIT-LDA calculated one ($B_0$= 85.86) the agreement between the calculated and the experimental values improves).}
\begin{ruledtabular}
\begin{tabular}{ccccc}
& &$\Gamma$ & L &	X	  \\ \hline
LO & theory & 1.12 & 1.39 & 1.22\\
LO & exp & 0.90 $^a$  & 1.0 $^a$ & 1.1 $^a$\\
TO& theory & 1.66& 1.71 & 1.69\\
TO& exp & 1.27 $^a$ & 1.0 $^a$ & -\\
LA& theory & - & 0.80 & 1.08\\
LA& exp & - & - & -\\
TA& theory & - & 1.35 & -1.09\\
TA& exp & 0.21; -0.90 $^b$ & -1.5 $^a$ & -1.2 $^a$\\
\end{tabular}

$^a$ Reference \onlinecite{Weinstein1977}

$^b$ Reference \onlinecite{Weinstein1984}
\end{ruledtabular}

\end{table*}

\section{Conclusions}\label{SecConclusions}
The original motivation of this work was to perform accurate measurements  of the specific heat of zb-ZnS over a broad temperature range, using samples of different isotopic compositions and to compare these with  \textit{ab initio} calculations based on phonon dispersion relations obtained from \textit{ab initio} band structures. Because of the availability of several band structure codes, we also decided to compare the results obtained using them with either LDA or GGA exchange-correlation Hamiltonians. The agreement between experimental and theoretical results has been found to be quite satisfactory, the former lying between the rather close LDA and GGA calculations. Because of the extensive computational tools we developed, we decided to use them to tackle a few extant problems involving the electronic structure and the lattice dynamics. We considered the effect of spin-orbit interaction on the lattice parameters and the lattice dynamics and found it to be negligible (contrary to the results obtained earlier for materials with heavier atoms, such as PbTe or Bi). We conjectured that the anomalously small \textit{s-o} splitting at the top of the valence bands of zb-ZnS was due to admixture of 3$d$-Zn core levels. In order to prove this conjecture, we performed calculations for ZnS in the centrosymmetric rock salt structure. In this structure no 3$d$-Zn admixture is possible at the top of the valence bands and the \textit{s-o} splitting has the larger expected value. We also calculated the pressure dependence of the \textit{s-o} splitting of zb-ZnS and found it to anomalously decrease with pressure, a fact that we have attributed to the pressure dependent contribution of the 3$d$-Zn electrons. Another effect of the 3$d$ admixture is reflected in the valence bands of rs-ZnS which, like those of similar rs-type materials (e.g. AgCl) have a maximum away from the center of the Brillouin zone (our Ref. \onlinecite{Bassani1965}. We have confirmed this by calculating the band structure of this phase\cite{Ves1990} which erroneously turned out to be semimetallic because of the "gap problem". By applying an ad hoc "scissors operator" we found that the indirect gap should be about 1.75 eV, in reasonable agreement with that found experimentally (1.9 eV, Ref. \onlinecite{Ves1990}). The zb phase of ZnS has linear splittings of the valence bands around the $\Gamma$ point of the BZ. We have calculated these splittings which should be of interest in the field of spintronics.

We have also investigated pressure effects on the crystal structure of ZnS. The zb-rs phase transition was calculated to occur at 15.7 GPa when using the LDA Hamiltonian, whereas the use of the GGA led to a transition pressure about 1 GPa higher, a fact which is probably related to the different lattice parameters found at zero pressure when using either of the two approximations. We have performed several studies of the lattice dynamics of zb-ZnS. The \textit{ab initio} calculations agree rather well with the experimental data obtained by INS and Raman spectroscopy. We have also presented calculation of the phonon density of states (including its projection on each of the constituent atoms) and the optical two-phonon DOS. The latter has been compared with second order Raman spectra.

Finally, we have calculated the pressure dependence of the phonon frequencies at the $\Gamma$, X and L high symmetry points and the corresponding Gr\"uneisen parameters. The Gr\"uneisen parameter of the transverse (Born) effective charge has also been evaluated.

\begin{acknowledgments}
A.H.R. has been supported by CONACyT Mexico
under project J-59853-F and by PROALMEX/DAAD.
Further computer resources have been provided by CGSTIC department at CINVESTAV-Mexico.
A. M. acknowledges the financial support from the Spanish MCYT under grants MAT2007-65990-C03-03,  CSD2007-00045 and the supercomputer  resources provides by the Red Espa\~{n}ola de Supercomputaci$\acute{\rm{o}}$n.
We are also indebted to XYZ for a critical reading of the manuscript.

\end{acknowledgments}

\end{document}